# Planck's Radiation Law in the Quantized Universe

Rainer Collier

Theoretisch-Physikalisches Institut, Friedrich-Schiller-Universität Jena

Max-Wien-Platz 1, 07743 Jena, Germany

E-mail: rainer@dr-collier.de

**Abstract.** Physical research looks for clues to quantum properties of the gravitational field. On the basis of the common Schrödinger theory, a simple model of the quantization of a Friedmann universe comprising dust and radiation is investigated. With regard to energy quantization, the result suggests a universal limitation of the energy spacing between neighbouring quantum states of $\Delta E \leq M_* c^2$ ($M_*$ Planck mass). Applied to black-body radiation, a modified Planck radiation law follows. If this could be verified in the laboratory, it would provide a direct hint at quantum properties of the space-time manifold.

## 1 Introduction

With increasing frequency, physical research is raising the question about experimental, astrophysical or convincing theoretical clues to quantum properties of the gravitational field. The present study attempts, via a rather simple form of quantizing the universe, to come to conclusions that can be verified in the laboratory. For this purpose, let us go the classical way of Schrödinger's quantization, deriving the Hamiltonian function of the universe from both Newton's and Einstein's gravitation theories. From the result of quantizing the Friedmann universe comprising dust and some radiation, there follows a limitation of the energy of radiation quanta by the Planck energy. If we take this quantum limitation into account when deducing the radiation laws for the electromagnetic black-body radiation, we arrive at corrected Planck radiation laws that may be verifiable in the laboratory. Thus, we could prove quantum properties of the space-time manifold in a direct test. Specifically, let us follow the procedure described below.

Recalling Schrödinger's theory of the electromagnetic Kepler problem, we know that the proper energy eigenvalues $E_n \sim 1/n^2$ appear already in the quantization of the radial zero angular momentum problem. Let us therefore take the classical energy conservation law of the corresponding gravitational Kepler problem and rewrite it in terms of the mass $M$ generating the gravitational field. The distance coordinate $r$ then appearing in the equation can also in this case be regarded as the configuration space variable of the associated quantization problem. Its physical meaning can be found by comparison with Einstein's gravitation theory. It is the scale parameter of the homogeneous and isotropic Friedmann-Lemaître-Robertson-Walker universes (FLRW universes). By finding the associated Lagrangian and Hamiltonian functions we can establish the corresponding Schrödinger equation quite formally. It is identical to the Wheeler-DeWitt equation for FLRW universes, provided we restrict the matter therein to negligibly weakly excited one (i.e., matter in the ground state). As our objective is only the determination of energy eigenvalues (and some discrete structures derived from them), we will not discuss here the various opinions on how to interpret the wave function in the quantized universe. Comments on the application of quantum theory to the universe as a whole, concerning Hamiltonian operator and wave function, boundary conditions, Wheeler-DeWitt equation, path integral formulation, loop quantum gravitation, etc., can be found in the literature [1] to [13] and summarized in [48].

For dust matter, our formulation yields the known energy quantization $E_n = \sqrt{2n+1}\, E_*$, in which $E_*$ denotes the Planck energy (or an energy unit in the vicinity of Planck energy). Subsequently we will refine the universe model by adding some excited electromagnetic radiation to the dust matter. This results in a mutual dependence of the two energy kinds in form of a new quantisation law $E_n = \sqrt{(2n+1)(1-\hbar\omega/E_*)}\, E_*$. The consequent limitation of the energy portions for electromagnetic radiation quanta can be generalized to the following universally valid hypothesis:

*As a result of the discrete micro-structure of space-time manifold, quantum jumps between the energy eigenstates of a quantum system are universally limited to $\Delta E \leq E_*$.*

This hypothesis should be added to the approved fundamental assumptions of quantum theory, until a satisfying theory of quantum gravitation has emerged that makes the hypothesis superfluous. One should be aware, though, that such a hypothesis is of importance only for quantum theory, as classical physics knows continuous energy changes only.

From the said „hypothesis of maximum quantum jump", we can immediately draw several conclusions that can be found in chapter 6 of this paper. This will also supply a universal cut-off parameter for some improper integrals of quantum field theory [37].

It is interesting to apply the said hypothesis to the partition function of black body radiation. It will yield a corrected Planck radiation law, with the consequence of a modified Stefan-Boltzmann law for the entire radiant energy and the radiation pressure. Finally, the modified thermodynamic functions of radiant energy $U(V,T)$, free energy $F(V,T)$, number of particles $N(V,T)$, entropy $S(V,T)$ and pressure $p(V,T)$ will be given, and their temperature behaviour for $T \to \infty$ will be discussed. The fluctuation of the energy about the mean value also contains an interesting additional term as compared to Planck's formula.
An application of the new state equation $p = p(V,T)$ to the radiation era in the very early cosmic evolution is obvious and will be realized in the next paper.

## 2 Finding the Hamiltonian function

To find a simple approach to the quantization of Friedmann universes on the basis of Schrödinger's quantum theory, let us first regard the case of the restricted, radial motion (zero orbital angular momentum) of a test mass $m$ in the gravitational field of a mass $M$ having the energy balance

$$\frac{m}{2}\dot{r}^2 - \frac{GmM}{r} = K, \qquad (2.1)$$

where $r$ is the distance between the two masses, $\dot{r} = dr/dt$ is their radial velocity, $K$ is the resulting total energy of motion, and $G$ is Newton's gravitational constant. By quantizing this special radial Kepler problem using the Schrödinger theory, we would already obtain the correct discretization $K_n$ of the energy eigenvalues:

$$K_n = -\frac{const}{n^2} \quad , \quad n = 1, 2, 3, \cdots. \quad (2.2)$$

However, in the conservation law (2.1) there also appears, in addition to the energy constant $K$, the mass constant $M$. If we solve (2.1) in terms of $M$

$$\frac{1}{G}\left[\frac{1}{2}r\dot{r}^2 - \frac{K}{m}r\right] = M \quad (2.3)$$

and introduce the abbreviations

$$E = Mc^2 \quad , \quad \frac{k}{2} = -\frac{K}{mc^2} \quad (2.4)$$

we obtain a conservation law of the form

$$\frac{c^2}{2G}\left[r\dot{r}^2 + kc^2 r\right] = E, \quad (2.5)$$

which expresses the time constancy of the total mass $M = E/c^2$. Associated with the conservation law (2.5) are the Hamiltonian and Lagrangian functions

$$H = \frac{c^2}{2G}\left[r\dot{r}^2 + kc^2 r\right] \quad , \quad L = \frac{c^2}{2G}\left[r\dot{r}^2 - kc^2 r\right] . \quad (2.6)$$

Now let us see which classical motion problem is described by this Lagrangian or Hamiltonian and what the associated quantum theory looks like.

## 3 The equation of motion of Friedmann universes

Let us identify the coordinate $r$ in (2.6) with the cosmological scale factor of the Robertson-Walker metric for homogeneous and isotropic space-times,

$$ds^2 = g_{\mu\nu}dx^\mu dx^\nu = c^2 dt^2 - r^2 d\ell^2 \quad , \quad (3.1)$$

$$d\ell^2 = \gamma_{\alpha\beta}dx^\alpha dx^\beta = \frac{d\xi^2}{1-k\xi^2} + \xi^2(d\vartheta^2 + \sin^2\vartheta d\varphi^2) \quad ,$$

where $d\ell^2$ denotes the line element of a three-dimensional position space of constant curvature. To compute the principal function $W$ of Einstein's theory of gravitation

$$W = \frac{1}{c}\int \mathcal{L}_G \, d^4x \quad , \quad \mathcal{L}_G = -\frac{1}{2\kappa}R\sqrt{-g} \, , \quad (3.2)$$

we determine the Lagrangian density $\mathcal{L}_G$ for the space-time structure (3.1). For the curvature scalar $R$ we obtain

$$R = \frac{6}{c^2}\left[\frac{\ddot{r}}{r} + \left(\frac{\dot{r}}{r}\right)^2 + \frac{c^2 k}{r^2}\right] . \quad (3.3)$$

With $(1/c)\sqrt{-g}\,d^4x = r^3\sqrt{\gamma}\,d^3x\,dt$ and $\ddot{r}r^2 = (\dot{r}r^2)^{\cdot} - 2r\dot{r}^2$, $W$ is given the form

$$W = -\frac{6}{2\kappa c^2}\int \sqrt{\gamma}\,d^3x \cdot \int\left[(\dot{r}r^2)^{\cdot} - r\dot{r}^2 + rc^2k\right]dt. \tag{3.4}$$

For the space integral we choose the value $4\pi/3$ and drop the term containing the total time derivative from the time integral. Inserting Einstein's gravitational constant $\kappa = 8\pi G/c^4$ yields

$$W = \frac{c^2}{2G}\int\left[r\dot{r}^2 - rkc^2\right]dt = \int L\,dt, \tag{3.5}$$

in which the Lagrangian function $L$ given in (2.6) appears. The equation of motion derived from the Lagrangian (2.6) reads

$$\left(\frac{\partial L}{\partial \dot{r}}\right)^{\cdot} - \frac{\partial L}{\partial r} = \frac{c^2}{2G}\left[2r\ddot{r} + \dot{r}^2 + kc^2\right] = 0 \tag{3.6}$$

and the associated Hamiltonian $H(p_r, r)$ is

$$H = p_r\dot{r} - L\,,\quad p_r = \frac{\partial L}{\partial \dot{r}} = \frac{c^2}{G}r\dot{r}, \tag{3.7}$$

$$H = \frac{G}{2c^2}\frac{p_r^2}{r} + \frac{c^4 k}{2G}r = \frac{c^2}{2G}\left[r\dot{r}^2 + rkc^2\right] = E = const, \tag{3.8}$$

as also stated in (2.6). $H = E = Mc^2$ is, at the same time, a first integral of the equation of motion (3.6). With

$$M = \frac{4\pi}{3}\rho r^3 = const \tag{3.9}$$

in (3.8), we recognize the Friedmann equation for a dust universe dominated by matter, having the rest mass density $\rho = \rho(t)$,

$$\dot{r}^2 + kc^2 = \frac{8\pi c^2}{3}r^2\rho. \tag{3.10}$$

Equations (3.6) and (3.8) are, respectively, the $(\alpha,\alpha)$- and $(0,0)$-components of Einstein's field equations $G_{\mu\nu} + \kappa T_{\mu\nu} = 0$ in the metric (3.1), as can be read, e.g., in [14] or [15]. The conservation law (3.9) is a consequence of the associated integrability condition $T^{\mu\nu}{}_{;\nu} = 0$.

This consideration makes it clear that the canonical functions $H$ and $L$ newly constructed from the conservation law (2.1) have an exact meaning not in Newton's, but in Einstein's gravitation theory: They are the Hamiltonian and Lagrangian functions for the gravitational field of a homogeneous and isotropic mass distribution of a dust universe dominated by matter. Thereby, the coordinate $r$ loses its original meaning as the distance between two masses $m$ and $M$. As a cosmological scale parameter it now describes, in the metric (3.1), the Gaussian curvature $k/r^2$ of the three-dimensional space slices $t = const.$.

# 4 Quantization of the Friedmann universe

Irrespective of the physical meaning of the coordinate $r$ one can formally quantize the motion problem that is based on the Lagrangian function $L = L(r,\dot{r})$ from (2.6). For this purpose we put $k = +1$ and, using (3.7), write the Hamiltonian $H$ belonging to $L$ in terms of the generalized coordinates $H = H(r, p_r)$,

$$H = \frac{c^2}{2G}\left[\frac{1}{r}\left(\frac{p_r}{\frac{c^2}{G}}\right)^2 + rc^2\right] = E . \qquad (4.1)$$

Introducing Planck units

$$L_* = \sqrt{\frac{\hbar G}{c^3}} \quad , \quad P_* = \sqrt{\frac{\hbar c^3}{G}} \quad , \quad M_* = \sqrt{\frac{\hbar c}{G}} \quad , \qquad (4.2)$$

with the relations

$$L_* P_* = \hbar \quad , \quad L_* \Omega_* = c \quad , \quad E_* = M_* c^2 \qquad (4.3)$$

and multiplying equation (4.1) by $r$, we can recast (4.1) as

$$\frac{p_r^2}{2M_*} + \frac{M_*}{2}\Omega_*^2 r^2 - \frac{E}{L_*}r = 0 \qquad (4.4)$$

By means of a quadratic supplement we obtain an energy equation that corresponds to a harmonic oscillator whose rest position is displaced from $r = 0$ to $r = b$:

$$\frac{p_r^2}{2M_*} + \frac{M_*}{2}\Omega_*^2 (r-b)^2 = \frac{E^2}{2E_*} \quad , \qquad (4.5)$$

$$b = L_* \frac{E}{E_*} = L_* \frac{M}{M_*} = \frac{GM}{c^2} = \frac{1}{2}R_s \quad . \qquad (4.6)$$

Equation (4.5) therefore describes an oscillator of a mass $M_*$ which, within the Schwarzschild radius $R_s = 2b$ of mass $M$, performs harmonic oscillations at the frequency $\Omega_*$ in the range of $0 \leq r \leq R_s$ about the rest position $r = R_s/2$.

Now, by the substitutions in (4.5) of

$$p_r \Rightarrow \frac{\hbar}{i}\frac{d}{dr} \quad , \quad r \Rightarrow r \quad , \quad \{r \in \mathbb{R}\} \qquad (4.7)$$

we can proceed to the quantum theory of the problem and obtain the following Schrödinger equation,

$$\frac{d^2\Phi}{dr^2} + \frac{2M_*}{\hbar^2}(\mathcal{E} - U)\Phi = 0 \quad , \qquad (4.8)$$

in which we have summarizingly set

$$\mathcal{E} = \frac{E^2}{2E_*} \quad , \quad U = \frac{1}{2} M_* \Omega_*^2 (r-b)^2 \quad . \tag{4.9}$$

As we can see, the notation (4.1) of the classical Hamiltonian already preselects a particular possible order of operators ([1], [3], [4], [7]) on transition from (4.1) to (4.8). Also, the quantum equation (4.8) corresponds to the Wheeler-DeWitt equation of quantum gravitation for the case that the universe is filled only by negligibly weakly excited matter. Therefore, the Schrödinger equation (4.8) is a version of describing the quantum nature of the Friedmann universe that is possible (with regard to the operator order selected) as well as simple (with regard to the assumed ground state structure of matter). Quantized FLRW universes with scalar fields as matter, with radiation and cosmological constant, are dealt with in [6], [10], [11], [17], [18], [19], [20], [23] and [24]. For investigations into the hermiticity of Hamiltonian operators, associated boundary conditions and Hilbert space structures, see [1], [5] and [16].

As the zero-point shifting in the potential (4.9) of the harmonic oscillator has no influence on its energy eigenvalues, we can immediately state the eigenvalues for the quantity $\mathcal{E}$:

$$\mathcal{E}_n = \left(n + \frac{1}{2}\right) \hbar \Omega_* = \left(n + \frac{1}{2}\right) \frac{\hbar c}{L_*} = \left(n + \frac{1}{2}\right) E_* \quad . \tag{4.10}$$

With the aid of the relations (4.2), (4.3) and (4.9), we get

$$E_n^2 = 2E_* \left(n + \frac{1}{2}\right) \hbar \Omega_* = (2n+1) E_*^2 \quad , \tag{4.11}$$

$$E_n = \sqrt{2n+1}\, E_* \quad , \quad M_n = \sqrt{2n+1}\, M_* \quad . \tag{4.12}$$

A similar result was first achieved by B. DeWitt [21] by means of WKB studies on a differently structured Hamiltonian operator (with another operator order).

Here, now, there are the associated eigenfunctions $\Phi_n$: The solution of (4.8) for the eigenvalue $E_n$ is

$$\Phi_n(r) = \left(\frac{1}{\sqrt{n! 2^n \sqrt{\pi} L_*}}\right) H_n\left(\frac{r-b_n}{L_*}\right) \exp\left[-\frac{1}{2}\left(\frac{r-b_n}{L_*}\right)^2\right] \quad . \tag{4.13}$$

There $H_n$ denotes the Hermitian polynomials, which, via $b_n$, are a function of $n$ also in the argument. Since the expectation value of $(r - b_n)$ is known to vanish,

$$\langle \Phi_n | r - b_n | \Phi_n \rangle = 0 \quad , \tag{4.14}$$

we get, for the expectation value of $r$, using (4.6) and (4.12),

$$r_n \equiv \langle \Phi_n | r | \Phi_n \rangle = b_n \quad , \quad b_n = L_* \frac{E_n}{E_*} = \sqrt{2n+1}\, L_* \quad , \quad r_n = \sqrt{2n+1}\, L_* \quad . \tag{4.15}$$

In summary it can be stated that the energy content and the scale factor of this quantized Friedmann universe are linked as

$$E_n = \sqrt{2n+1}\, E_* \quad , \quad r_n = \sqrt{2n+1}\, L_* \quad , \quad n = 0,1,2,\cdots \tag{4.16}$$

The ground state $n = 0$ of the quantum universe having the least energy and the smallest scale factor is given by the Planck values $E_0 = E_*$, $r_0 = L_*$. According to (4.16) there is no vanishing expectation value for the scale factor $r$.

## 5 Refinement of the universe model

We can refine the universe model by permitting not only the dust of density $\rho_m$ but also radiation contributions of density $\rho_s$. Assume that both contributions do not substantially interact with one another. The energy-momentum tensor now has the form

$$T_{\mu\nu} = \left( \rho + \frac{p}{c^2} \right) u_\mu u_\nu - p\, g_{\mu\nu} \quad , \tag{5.1}$$

$$\rho = \rho_m + \rho_s \quad , \qquad p = p_m + p_s \tag{5.2}$$

with the equations of state

$$p_m = 0 \quad , \quad p_s = \frac{1}{3}\rho_s c^2 \quad . \tag{5.3}$$

In the metric (3.1), the Einstein equations reduce to

$$\dot{r}^2 + kc^2 = \frac{8\pi G}{3} r^2 \rho \qquad \longrightarrow \qquad (0,0)\text{-component}, \tag{5.4}$$

$$2r\ddot{r} + \dot{r}^2 + kc^2 = -8\pi G\, r^2 \frac{p}{c^2} \qquad \longrightarrow \qquad (\alpha,\alpha)\text{-components}. \tag{5.5}$$

The integrability condition $T^{\mu\nu}{}_{;\nu} = 0$ leads to the thermodynamic relation

$$\left[ (\rho c^2) V \right]^{\cdot} + p \dot{V} = 0 \quad , \tag{5.6}$$

which, with the equations of state (5.3) and the definitions

$$V = \frac{4\pi}{3} r^3 \quad , \quad M_m = \rho_m V \quad , \quad M_s = \rho_s V \tag{5.7}$$

can also be given the form

$$\dot{M}_m + \frac{1}{r}(M_s r)^{\cdot} = 0 \quad . \tag{5.8}$$

Because of the assumed absence of interaction of dust and radiation, we satisfy (5.8) by the two separate settings

$$M_m = const = M \quad , \quad M_s r = const = \mu\, L_* \quad . \tag{5.9}$$

In this set of field equations (5.4), (5.5) and (5.6), too, the ($\alpha,\alpha$)-equation can be derived from the (0,0)-equation (5.5) by means of time differentiation and inclusion of the integrability condition (5.6). If the conditions (5.9) are satisfied, the ($0,0$)-equation (5.4) alone describes the dynamic nature of the universe. Inserting these conditions (5.9) into (5.4), we get a balance of the form (Friedmann equation)

$$\dot{r}^2 + kc^2 = \frac{8\pi G}{3} r^2 (\rho_m + \rho_s) \quad , \tag{5.10}$$

or also, with (5.7),

$$r(\dot{r}^2 + kc^2) = 2G\left(M + \frac{\mu L_*}{r}\right) \quad . \tag{5.11}$$

With $E = Mc^2$ and $\varepsilon = \mu c^2$, this can be converted into an energy conservation equation of the form

$$\frac{c^2}{2G}\left[r(\dot{r}^2 + kc^2)\right] - \frac{\varepsilon L_*}{r} = E = const \tag{5.12}$$

Now we can interpret this equation as the energy conservation relation of a dynamic problem with the Hamiltonian

$$H = \frac{c^2}{2G}\left[r(\dot{r}^2 + kc^2)\right] - \frac{\varepsilon L_*}{r} \quad . \tag{5.13}$$

The associated Lagrangian then reads

$$L = \frac{c^2}{2G}\left[r(\dot{r}^2 - kc^2)\right] + \frac{\varepsilon L_*}{r} \quad . \tag{5.14}$$

We can check that, taking into account the conditions (5.9), the Euler-Lagrange equations of motion with $L$ from (5.14) are just the ($\alpha,\alpha$)-Einstein equations (5.5) and that the associated energy balance $H = E = const$ from (5.12) constitutes the-($0,0$) Einstein equation (5.4). Inserting the canonical momentum resulting from (5.14)

$$p_r = \frac{\partial L}{\partial \dot{r}} = \frac{c^2}{G} r \dot{r} \tag{5.15}$$

into (5.12) and using the Planck units (4.2), (4.3), we have

$$\frac{p_r^2}{2M_*} + \frac{M_*}{2}\Omega_*^2 r^2 - \frac{E}{L_*} r - \varepsilon = 0 \quad . \tag{5.16}$$

Here again, we can find, by a quadratic supplement, the equation of motion of a harmonic oscillator in which the zero-point shift $R$ retains the earlier value (4.6)

$$\frac{p_r^2}{2M_*} + \frac{M_*}{2}\Omega_*^2 (r-b)^2 = \frac{E^2}{2E_*} + \varepsilon \quad . \tag{5.17}$$

The replacements (4.7) again get us to the Schrödinger equation (Wheeler-DeWitt equation)

$$\frac{d^2\Phi}{dr^2} + \frac{2M_*}{\hbar^2}(\varepsilon - U)\Phi = 0 \quad , \tag{5.18}$$

in which we have now summarizing put

$$\mathcal{E} = \frac{E^2}{2E_*} + \varepsilon \quad , \quad U = \frac{1}{2} M_* \Omega_*^2 (r-b)^2 \quad . \tag{5.19}$$

The eigenvalues of the quantity $\mathcal{E}$ can be stated immediately

$$\mathcal{E}_n = \left(n + \frac{1}{2}\right)\hbar\Omega_* = \left(n + \frac{1}{2}\right)E_* \quad . \tag{5.20}$$

With (5.19) we then get, for the energy constant $E$

$$E_n = \sqrt{\left[(2n+1)E_* - 2\varepsilon_n\right]E_*} \quad . \tag{5.21}$$

This result allows us to draw some interesting conclusions:
With the sole existence of matter in dust form ($\varepsilon_n = 0$), the earlier energy eigenvalues reappear,

$$E_n = \sqrt{2n+1}\, E_* \quad . \tag{5.22}$$

With the sole existence of matter in radiation form ($E_n = 0$), there results an energy quantization independent of its spectral structure, of the form

$$\varepsilon_n = \left(n + \frac{1}{2}\right)E_* \quad . \tag{5.23}$$

Similar results as those for dust in (5.22) and for radiation in (5.23) have already been achieved by Padmanabhan [22] using the method of conform quantization.
With the simultaneous existence of dust and radiation, however, according to (5.21), the energy eigenvalues of the two matter configurations influence each other in such a way that the occurrence of quantized radiation causes a decrease in the energy levels of the matter contribution.

## 6 The maximum energy gap hypothesis

Let us apply these findings to a situation in which there is, in addition to dust matter, some radiation that is an ensemble of harmonic oscillators of a single angular frequency $\omega$. These may be distributed among the energy levels

$$\varepsilon_n = \left(n + \frac{1}{2}\right)\hbar\omega \tag{6.1}$$

according to common quantum theory. For the matter contribution, then, (5.21) yields the

quantum condition

$$E_n = \sqrt{(2n+1)\left(1 - \frac{\hbar\omega}{E_*}\right)} E_* \quad . \tag{6.2}$$

For the radiation contribution, it follows that the energy quanta of the radiation field must be limited in magnitude by the Planck energy,

$$\hbar\omega \leq E_* \quad . \tag{6.3}$$

This also means, however, that the magnitude of the energy spacing between neighbouring energy eigenstates of harmonic oscillators is limited. This result, obtained for a special quantum system, can be extended into a generally valid statement, which is obviously a consequence of a discontinuous structure of the space-time manifold:

*The magnitude of the energy spacing $\Delta E$ between neighbouring energy eigenstates of a quantum system is universally limited by the Planck energy $E_* = M_* c^2$,*

$$\Delta E \leq E_* \quad . \tag{6.4}$$

It should be pointed out that even quantum objects as large as the quantum black holes described in [38], [39] and [40] obviously satisfy such a condition. So, it was pointed out in [44] – [47] that quantisation of Schwarzschild black holes leads to the same Schrödinger equation from harmonic oszillator type wih mass quantization $M_n^2 = (2n+1)M_*^2$ as our quantized universe with the Schrödinger equation (4.8) and the quantum condition (4.12).

However, as we know that transitions in a harmonic oscillator can only take place between neighbouring energy levels, we can extend the statement (6.4) into the following hypothesis:

*Maximum quantum jump hypothesis*

*The magnitude of quantum jumps $\Delta E$ between the energy eigenstates of a quantum system is limited by the Planck energy $E_* = M_* c^2$,*

$$\Delta E \leq E_* \quad . \tag{6.5}$$

This hypothesis lends the Planck energy $E_*$ a fundamental importance in the whole quantum physics. We note that the exact magnitude of $E_* = M_* c^2 = \alpha M_{Pl} c^2$ ($M_{Pl} = \sqrt{\hbar c / G}$, Planck-Mass) should only be ascertained by experiment or observation [41], with $\alpha$ certainly to have unity order of magnitude. For the time being, though, let us continue to reckon with $\alpha = 1$.

Some conclusions can be drawn from this hypothesis straightaway. For this purpose, let us regard the creation of a free elementary particle of rest mass $m_0$ and magnitude of momentum $\Delta p$. Its excitation energy $\Delta E$, because of (6.5), satisfy the relation

$$(\Delta E)^2 = (c\Delta p)^2 + (m_0 c^2)^2 \leq E_*^2 \quad . \tag{6.6}$$

In case of a vanishing momentum, then, the Planck mass $M_*$ is an upper bound to the rest mass spectrum of elementary particles,

$$m_0 \leq \frac{E_*}{c^2} = M_* \quad . \tag{6.7}$$

In the case vanishing rest masses also the magnitude of momentum transfers is bounded by the Planck momentum,

$$\Delta p \leq \frac{E_*}{c} = P_* \quad . \tag{6.8}$$

For photons ($m_0 = 0$), bounds to energy and momentum immediately result from (6.6),

$$E = \hbar\omega \leq E_* \quad , \quad p = \frac{\hbar\omega}{c} \leq P_* \quad . \tag{6.9}$$

If we interpret $\Delta p$ as momentum uncertainty, there exists, with maximum uncertainty $\Delta p = P_*$, a fundamental minimum position uncertainty $\Delta q$. With $\Delta p \Delta q \geq \hbar/2$, there follows

$$\Delta q \geq \frac{1}{2} L_* \quad . \tag{6.10}$$

The statements (6.7), (6.8), (6.10) are the subject of many investigations, considerations and presumptions, which are summarized, e.g., in [25] and [49] – [58].

At the end of this section it seems appropriate to note that at present there exists no quantum theory with a universal energy gap limitation $\Delta E \leq E_*$. As, according to (4.2), (4.3), the gravitational constant $G$ appears in $E_*$ in addition to $c$ and $\hbar$, a satisfactory theory of quantum gravitation might possibly support the hypothesis (6.5).

## 7 Application to the photon gas

Here we use Einstein's proven model of cavity radiation [26]. For this purpose, we regard electromagnetic radiation that is contained in the volume $V$ and has the spectral energy density $\bar{u}(\omega,T)$, the frequency $\omega$ and the temperature $T$. Let the volume $V$ also contain microsystems (atoms) which can only adopt two states of the energies $E_1$, $E_2$ ($E_2 > E_1$, laser model). Between the electromagnetic radiation and the microsystems, there will happen (per unit time interval) $Z_{12}^{abs}$ absorption processes, $Z_{21}^{spn}$ spontaneous emission processes and $Z_{21}^{ind}$ induced emission processes.

As we know,

$$Z_{12}^{abs} = B_{12} \cdot N_1 \cdot \bar{u} \quad , \quad Z_{21}^{spn} = A \cdot N_2 \quad , \quad Z_{21}^{ind} = B_{21} \cdot N_2 \cdot \bar{u} \quad , \tag{7.1}$$

where $B_{12}$, $A$, $B_{21}$ are proportionality factors and $N_1$, $N_2$ are the numbers of atoms in the ground state $E_1$ and excited state $E_2$, respectively.

In a thermodynamic equilibrium, necessarily,

$$Z_{12}^{abs} = Z_{21}^{spn} + Z_{21}^{ind} \quad . \tag{7.2}$$

The Boltzmann distribution yields the number $N_2$ of microsystems when excited at the temperature $T$:

$$N_2 = N_1 \exp\left(-\frac{\Delta E}{k_B T}\right) = N_1 \exp\left(-\frac{\hbar\omega}{k_B T}\right) \quad , \tag{7.3}$$

in which we have put $\Delta E = E_2 - E_1 = \hbar\omega$. With (7.1) and (7.3) substituted in (7.2), the resulting spectral energy density $\bar{u}$ is given by

$$\bar{u}(\omega,T) = \frac{A}{B_{12}} \cdot \frac{1}{\exp\left(\dfrac{\hbar\omega}{k_B T}\right) - \dfrac{B_{21}}{B_{12}}} \quad . \tag{7.4}$$

The factors $B_{12}, A, B_{21}$ can be set (at a fixed $\hbar\omega$) by two requirements on temperature behaviour:

First requirement (Wien's law)

For sufficiently low temperatures, let Wien's law $\bar{u}_w = \bar{u}_w(\omega,T)$ apply. Then (7.4) becomes

$$\lim_{T \to 0} \bar{u} = \bar{u}_w \quad , \quad \bar{u}_w = \frac{\omega^2}{\pi^2 c^3}(\hbar\omega)\exp\left(-\frac{\hbar\omega}{k_B T}\right) \quad \to \quad A = \frac{\omega^2}{\pi^2 c^3}(\hbar\omega) B_{12} \quad . \tag{7.5}$$

Thus, at the temperature $T$, the spectral energy density $\bar{u}$ and the mean energy $\bar{\varepsilon}$ at the level $\hbar\omega$ adopt the following form:

$$\bar{u} = \frac{\omega^2}{\pi^2 c^3}\bar{\varepsilon} \quad , \quad \bar{\varepsilon} = \frac{\hbar\omega}{\exp\left(\dfrac{\hbar\omega}{k_B T}\right) - \dfrac{B_{21}}{B_{12}}} \quad . \tag{7.6}$$

Second requirement (hypothesis):

For sufficiently high temperatures, due to the hypothesis (6.5) and the energy limit (6.9) resulting from it, quantum jumps between the energy levels $E_1$ and $E_2$ can take place up to a maximum of $\Delta E \leq E_*$ only. Therefore, for $T \to \infty$, necessarily

$$\lim_{T \to \infty} \bar{\varepsilon} = E_* \quad , \quad B_{21} = \left(1 - \frac{\hbar\omega}{E_*}\right) B_{12} \quad . \tag{7.7}$$

Now, the Einstein coefficient of induced emission becomes a function of the energy of the photons and vanishes when the exchanged photons reach the threshold energy $E_*$.

For the mean energy $\bar{\varepsilon}$ there results a corrected Planck distribution of the form

$$\bar{\varepsilon}(\omega, T) = \frac{\hbar\omega}{\exp\left(\dfrac{\hbar\omega}{k_B T}\right) - \left(1 - \dfrac{\hbar\omega}{E_*}\right)} \quad . \tag{7.8}$$

Some limiting cases of this formula are remarkable:

    For $E_* \to \infty$ (with $\omega, T$ optional), we have Planck distribution $\bar{\varepsilon} = \bar{\varepsilon}_{Pl}$.

    For $\hbar\omega \to E_*$ (with $T$ optional), we have Boltzmann distribution.

    For $T \to \infty$ (with $\omega$ optional), we have an energy limitation $\bar{\varepsilon} \to E_*$.

Other exact notations of the mean energy $\bar{\varepsilon}$ are of interest,

$$\bar{\varepsilon} = \frac{\bar{\varepsilon}_{Pl}}{1 + \dfrac{\bar{\varepsilon}_{Pl}}{E_*}} \quad , \qquad \frac{1}{\bar{\varepsilon}} = \frac{1}{\bar{\varepsilon}_{Pl}} + \frac{1}{E_*} \quad , \qquad \bar{\varepsilon}_{Pl} = \frac{\hbar\omega}{\exp\left(\dfrac{\hbar\omega}{k_B T}\right) - 1} \quad . \tag{7.9}$$

    (corrected Planck distribution)      (harmonic mean)      (Planck distribution)

By the way, the new radiation formula (7.8) contains all the four universal constants of nature: the velocity of light in vacuum $c$, the Planck constant $\hbar$, Newton's gravitational constant $G$ and the Boltzmann constant $k_B$.

## 8 Discussion of the new photon distribution

Let us investigate the new curve of spectral energy density $\bar{u}$ with the corrected Planck distribution for $\bar{\varepsilon}$ from (7.8):

$$\bar{u}(\omega, T) = \frac{\omega^2}{\pi^2 c^3} \bar{\varepsilon}(\omega, T) = \frac{1}{\pi^2 \hbar^2 c^3} \cdot \frac{(\hbar\omega)^3}{\exp\left(\dfrac{\hbar\omega}{k_B T}\right) - \left(1 - \dfrac{\hbar\omega}{E_*}\right)} \quad . \tag{8.1}$$

For this purpose, we introduce new dimensionless variables $z$ and $\theta$ in (7.8) and (8.1),

$$z = \frac{\hbar\omega}{k_B T} \quad , \quad \theta = \frac{k_B T}{E_*} \quad , \quad \theta z = \frac{\hbar\omega}{E_*} \quad , \tag{8.2}$$

$$\bar{\varepsilon} = (k_B T) \cdot \left\{ \frac{z}{\exp(z) - (1 - \theta z)} \right\} \quad , \quad \bar{u}(z,\theta) = \frac{(k_B T)^3}{\pi^2 \hbar^2 c^3} \cdot \left\{ \frac{z^3}{\exp(z) - (1 - \theta z)} \right\} \quad . \tag{8.3}$$

Figures 1 and 2 in the appendix show the function behaviour of $\bar{\varepsilon}(z,\theta)$ and $\bar{u}(z,\theta)$ for several parameter values $\theta$. The Planck laws of radiation result by virtue of $E_* \to \infty$, i.e. by virtue of $\theta \to 0$ within the braces (please note $E_*\theta = k_B T$). The maximum of the function $\bar{u} = \bar{u}(z,\theta)$ with $(\partial \bar{u}/\partial z) = 0$ results from the zero $z = z_0$ of the function $Q(z,\theta) = 0$,

$$Q(z,\theta) = (3-z)\exp z - (3 - 2\theta z) \quad , \quad Q(z_0,\theta) = 0 \quad , \tag{8.4}$$

in which the term containing the parameter $\theta$ indicates the change compared to Planck distribution, as can also be seen from figure 3 in the appendix.

The following boundary cases can be evaluated directly ($\theta = 0$: Planck case):

$$\begin{aligned} z_0 = z_{Pl} &= 2.821 \quad , & \theta &= 0 \quad (E_* \to \infty) \quad , \\ z_0 &= z_{Pl}\left(1 + 0.41 \frac{k_B T}{E_*}\right) \quad , & \theta &\ll 1 \quad (k_B T \ll E_*) \quad , \\ z_0 &= 3 \quad , & \theta &= \frac{1}{2} \quad \left(k_B T = \frac{1}{2} E_*\right) \quad , \\ z_0 &\approx \ln\left(2\frac{k_B T}{E_*}\right) \quad , & \theta &\gg 1 \quad (k_B T \gg E_*) \quad . \end{aligned} \tag{8.5}$$

In general, Wien's law of the displacement of the distribution maximum now reads

$$(\hbar\omega)_{max} = z_0(\theta) \cdot k_B T \quad . \tag{8.6}$$

Due to the temperature dependence of $z_0 = z_0(\theta)$, as can be calculated from (8.4), the maximum shifts (compared to the known Wien´s displacement law) with increasing temperature in the direction of increasing frequencies. Some examples are computed in (8.5). This shift is very small, until temperatures come close to the Planck temperature $k_B T = E_*$. If, e.g., $k_B T = E_*/100$, there follows $z_0 = 1.0041 \cdot z_{Pl}$. This causes a increase from $z_0 = z_{Pl} = 2.821$ to $z_0 = 2.833$.

Now we calculate the thermodynamic potentials $U(T,V)$, $F(T,V)$, $N(T,V)\cdots$. For this it is expedient to introduce the Planck volume $V_*$, the Planck density $\rho_*$ and the number of particles $N_*$,

$$V_* = 2\pi^2 L_*^3 \quad , \quad \rho_* = M_* / V_* \quad , \quad N_* = V / V_* \quad . \tag{8.7}$$

From (8.1) we get $U(T,V)$, taking the limiting energy $E_* = \hbar\Omega_* = M_* c^2$ and the spin degeneration factor $g_s = 2$ into account,

$$U(T,V) = V \cdot \int_0^{\Omega_*} \bar{u}(\omega,T)\, d\omega = \frac{g_s \cdot V}{2\pi^2 (\hbar c)^3} \int_0^{E_*} \frac{(\hbar\omega)^3\, d(\hbar\omega)}{\exp\left(\dfrac{\hbar\omega}{k_B T}\right) - \left(1 - \dfrac{\hbar\omega}{E_*}\right)}.$$

With the substitutions of (8.2) and (8.7) we get

$$U(\theta,V) = g_s N_* E_* \theta^4 \cdot \left\{ \int_0^{1/\theta} \frac{z^3\, dz}{\exp z - (1-\theta z)} \right\} \quad . \tag{8.8}$$

The free energy $F(T,V)$ has the form

$$F(T,V) = \frac{g_s V}{2\pi^2 (\hbar c)^3} (k_B T) \int_0^{E_*} \frac{(\hbar\omega)^2}{\left(1 - \dfrac{\hbar\omega}{E_*}\right)} \ln\left[1 - \left(1 - \frac{\hbar\omega}{E_*}\right) \exp\left(-\frac{\hbar\omega}{k_B T}\right)\right] d(\hbar\omega)$$

or also, with (8.2), (8.7)

$$F(\theta,V) = g_s N_* E_* \theta^4 \cdot \left\{ \int_0^{1/\theta} \frac{z^2}{1-\theta z} \ln\left[1 - (1-\theta z)\exp(-z)\right] \right\} dz \quad . \tag{8.9}$$

The number of particles $N(T,V)$ is calculated from (8.1) by means of the spectral particle density $\bar{v} = \bar{u}(\omega,T)/\hbar\omega,$ :

$$N(T,V) = V \cdot \int_0^{\Omega_*} \bar{v}(\omega,T)\, d\omega = \frac{g_s \cdot V}{2\pi^2 (\hbar c)^3} \int_0^{E_*} \frac{(\hbar\omega)^2\, d(\hbar\omega)}{\exp\left(\dfrac{\hbar\omega}{k_B T}\right) - \left(1 - \dfrac{\hbar\omega}{E_*}\right)},$$

and with (8.2) we have

$$N(\theta,V) = g_s N_* \theta^3 \cdot \left\{ \int_0^{1/\theta} \frac{z^2\, dz}{\exp z - (1-\theta z)} \right\} \quad . \tag{8.10}$$

With $F$ from (8.9) we also obtain the entropy $S$ and the pressure $p$ by

$$S = -\left(\frac{\partial F}{\partial T}\right)_V = kN_*\frac{\partial}{\partial \theta}\left[\frac{-F}{N_*E_*}\right] \quad , \quad p = -\left(\frac{\partial F}{\partial V}\right)_T = -\frac{F}{V} = \rho_*c^2\left[\frac{-F}{N_*E_*}\right] \tag{8.11}$$

The functions $U, F, N, S$ and the pressure $p$ can be seen in figures 4 to 8. The dependence on $V$ can frequently be found in the definition of $N_* = V/V_*$. Planck behaviour is achieved by $E_* \to \infty$ (with the definitions (8.2) taken into account). Marked deviations from Planck behaviour only occur close to the Planck energy, from approximately $k_BT \sim E_*/10$. This is shown particularly clearly by the run of the curve for the $(pV/U)$ ratio in figure 10. The form of the specific heat $C_V = (\partial U/\partial T)_V$ in figure 9 is also interesting. In the range around $k_BT \sim E_*/2$ it has a "hump", which is similar to the $C_V$ curve of a two-level system with an excitation gap of $\Delta = E_*$.

Below we give some approximate formulae for small and large $\theta = k_BT/E_*$, in which $\zeta(x)$ is the Riemann zeta function.

For $k_BT \ll E_*$ ($\theta \to 0$), there always appear the respective Planck functions $U_{Pl}, N_{Pl}, \cdots$ with a minor correction,

$$U(T,V) = U_{Pl}\left(1 - \eta_{(4)}\frac{k_BT}{E_*}\right) \quad , \quad U_{Pl} = 2N_*E_*\theta^4 I_\infty \quad , \quad I_\infty = \Gamma(4)\zeta(4) ,$$

(8.12)

$$\eta_{(4)} = 4\left(1 - \frac{\zeta(5)}{\zeta(4)}\right) \approx 0.17 \quad , \quad U_{Pl} = V\frac{\Gamma(4)\zeta(4)}{\pi^2}\left(\frac{k_BT}{\hbar c}\right)^3(k_BT) ,$$

$$N(T,V) = N_{Pl}\left(1 - \eta_{(3)}\frac{k_BT}{E_*}\right) \quad , \quad N_{Pl} = 2N_*\theta^3 J_\infty \quad , \quad J_\infty = \Gamma(3)\zeta(3) ,$$

(8.13)

$$\eta_{(3)} = 3\left(1 - \frac{\zeta(4)}{\zeta(3)}\right) \approx 0.30 \quad , \quad N_{Pl} = V\frac{\Gamma(3)\zeta(3)}{\pi^2}\left(\frac{k_BT}{\hbar c}\right)^3 .$$

$$F(T,V) = F_{Pl}\left(1 - \tfrac{3}{4}\eta_{(4)}\frac{k_BT}{E_*}\right) \quad , \quad F_{Pl} = -p_{Pl}V = -\tfrac{1}{3}U_{Pl} \tag{8.14}$$

$$S = -\left(\frac{\partial F}{\partial T}\right)_V = S_{Pl}\left(1 - \tfrac{15}{16}\eta_{(4)}\frac{k_BT}{E_*}\right) \quad , \quad S_{Pl} = \tfrac{1}{3}\left(\frac{\partial U_{Pl}}{\partial T}\right)_V . \tag{8.15}$$

For $k_B T \gg E_*$ ($\theta \to \infty$), some thermodynamic potentials now have finite limits. The respective approximations are as follows:

$$U(T,V) = \frac{2}{3} N_* E_* \left(1 - \frac{E_*}{k_B T}\right) , \tag{8.16}$$

$$N(T,V) = N_* \left(1 - \frac{E_*}{k_B T}\right) , \tag{8.17}$$

$$F(T,V) = 2 N_* E_* \left(\frac{1}{3} - C_\infty \cdot \left(\frac{k_B T}{E_*}\right)\right) \approx -2 C_\infty N_* \cdot (k_B T) = -TS \tag{8.18}$$

$$S(T,V) = 2 C_\infty N_* \cdot k_B \quad , \quad C_\infty = \frac{\pi^2}{6} - \frac{5}{4} \approx 0.395 \quad . \tag{8.19}$$

The relationships $F = U - TS$ and $pV = -F$ are satisfied in the approximations (8.12) to (8.19). At the limit $T \to \infty$, the relations

$$U_\infty = \frac{2}{3} N_* E_* \quad , \quad N_\infty = N_* \quad , \quad S_\infty = \tilde{N}_* \cdot k_B \quad , \tag{8.20}$$

appear, in which we have summarily put $\tilde{N}_* = 2 C_\infty N_*$. The energy or entropy in this case is approximately equal to the number of Planck volumes $V_*$ going into a given volume $V$, multiplied by the Planck energy $E_*$ or Boltzmann constant $k_B$, respectively. The number of photons is just equal to the number of Planck volumes $V_*$ filling the given volume $V$.

The state equation of such a matter configuration is obviously that of an ideal, non-interacting gas of $\tilde{N}_*$ Planck masses at an (extremely high) temperature $T$, as can be read from (8.18),

$$pV = -F = U_\infty - TS_\infty \approx S_\infty T = \tilde{N}_* k_B T \quad . \tag{8.21}$$

$$pV = \tilde{N}_* k_B T \tag{8.22}$$

The stress energy $pV$ now is no longer essentially determined by the internal energy $U$ but by the entropy $S$ dominating at these temperatures.

Also note that in the boundary case $T \to \infty$, there appears, by means of Boltzmann's entropy equation $S = k_B \cdot \ln W$ in the volume $V$ a maximum state number $W_\infty$,

$$S_\infty = k_B \cdot \ln W_\infty = k_B \cdot 2 C_\infty N_* = \tilde{N}_* \cdot k_B \sim \frac{V}{V_*} k_B \tag{8.23}$$

For the photon states $W_\infty$ in the Volume $V$ results

$$W_\infty = \exp\left(\frac{S_\infty}{k_B}\right) = [\exp(2C_\infty)]^{N_*} = (2.203)^{N_*} . \qquad (8.24)$$

Incidentally, for the smallest possible volume $V = V_*$ is by definition (8.7) $N_* = 1$ and therefor the Boltzmann constant $k_B$ is the smallest entropy portion (see 8.23).

## 9 Fluctuations

An interesting result is obtained when we investigate the fluctuation of the energy $\varepsilon$ about the mean value $\bar{\varepsilon}$ at the $\hbar\omega$ level. With (7.8) and $\beta = 1/k_B T$ it can be easily calculated with the known formula

$$\overline{(\Delta\varepsilon)^2} = \overline{\varepsilon^2} - \bar{\varepsilon}^2 = -\frac{\partial \bar{\varepsilon}}{\partial \beta} = \bar{\varepsilon}^2 \exp(-\beta\hbar\omega) . \qquad (9.1)$$

Once more, with the aid of (7.8), we express the exponential function by $\bar{\varepsilon}$ again, there follows for the relative square fluctuation

$$\overline{\left(\frac{\Delta\varepsilon}{\bar{\varepsilon}}\right)^2} = \frac{1}{\bar{\varepsilon}}\left[\bar{\varepsilon} + (\hbar\omega) - \frac{\bar{\varepsilon}\cdot(\hbar\omega)}{E_*}\right] , \qquad (9.2)$$

or also, using the mean occupation number $\bar{n} = \bar{\varepsilon}/\hbar\omega$:

$$\overline{\left(\frac{\Delta\varepsilon}{\bar{\varepsilon}}\right)^2} = \overline{\left(\frac{\Delta n}{\bar{n}}\right)^2} = \left[1 + \frac{1}{\bar{n}}\left(1 - \frac{\bar{\varepsilon}}{E_*}\right)\right] = \left[\left(1 - \frac{\hbar\omega}{E_*}\right) + \frac{1}{\bar{n}}\right] . \qquad (9.3)$$

The first two terms in (9.2) are already known from Planck's radiation formula. The first term describes the (classical) wave share, while the second one describes the (quantum-theoretical) particle share of the fluctuation. The additional third term obviously describes a kind of interaction between the other two: If (with fixed $\hbar\omega$) the classical share $\bar{\varepsilon}$ is increased to $\bar{\varepsilon} \to E_*$ by raising the temperature $k_B T \to \infty$, the quantum-theoretical share vanishes more and more. If vice versa (at a fixed temperature $k_B T$) the quantum-theoretical share $\hbar\omega$ is increased by raising the frequency $\hbar\omega \to E_*$, the classical share is vanishing gradually. This behaviour can also be readily seen from the notations (9.3) of the fluctuation formula: Depending on how we approach the limit energy $E_*$, the transition between classical and quantum-theoretical behaviour will become "soft" (look at [43]).

## 10 Concluding remarks

We have regarded a quantized Friedmann universe containing massive matter in the ground state and an excited monochromatic photon field. The result is a quantization of matter and radiation field in mutual dependence, so that a limitation of the energy spacing between neighbouring energy eigenstates of the radiation oscillators is formed. Generalizing this finding, we have proposed a hypothesis that quantum jumps between the energy eigenstates of a quantum system may universally not be larger than the Planck energy: $\Delta E \leq E_*$. The Planck energy $E_* = M_* c^2$ is thus given a fundamental importance throughout quantum physics. At the same time, because of its combination from the universal constants $c, \hbar$ and $G$, it suggests a connection with the outstanding theory of quantum gravitation.

If the said hypothesis is applied to the quantized electromagnetic radiation field, we receive a corrected Planck radiation law. As a consequence of this, the curves of the thermodynamic potentials $U, S, F, N$ and $p$ for very high temperatures $T \to \infty$ change considerably. Already at temperatures near the Planck temperature $k_B T \approx E_*$, there are distinct deviations from the common Planck radiation law. If one of these curves could be confirmed by laboratory tests or astrophysical observations, this would prove a direct connection between gravitational and quantum physics, and the role of the large Planck mass $M_* = \sqrt{\hbar c/G}$ would be better understood [42].

As in the vicinity of the big bang we would have to expect temperatures of the magnitude of the Planck temperature $k_B T = E_*$ or more, a subsequent study will apply these corrected state functions of the photon gas to investigating the behaviour of the early universe in the radiation era. Whether, e.g., the Stefan-Boltzmann law, on which the state equation $p = \rho c^2 / 3$ is based, can be applied to the extreme density conditions of the radiation field in the vicinity of the big bang singularity, is at least problematic, according to Schmutzer [27]. Papers by other authors (Refs. [28] to [36]) have also raised the question whether either fundamental physics (theory of special or general relativity, quantum theory, thermodynamics) or the equations of state need to be modified where physical research is dealing with energy densities in the Planck range.

The current views about existence, detection and theoretical characterisation of a universal lower bound for length measurements you can follow in the publications [49] to [58]. A current report on experimental verification of Lorentz invariance, fundamental to quantum and gravitational theory, can be found in [59].

**Appendix**

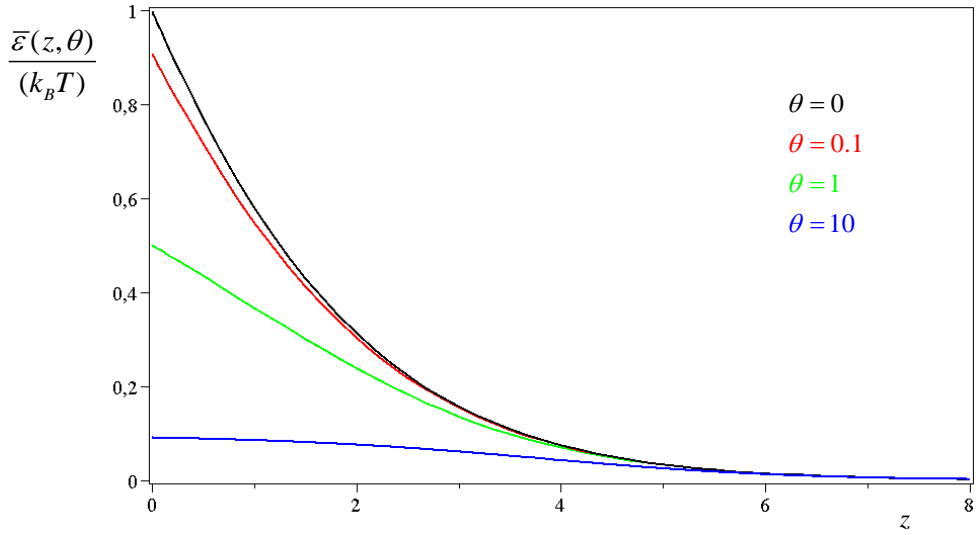

Fig. 1: Distribution of mean energy $\bar{\varepsilon}(z,\theta)$ among the photon frequencies $z = \hbar\omega / k_B T$ as a function of the temperature parameter $\theta = k_B T / E_*$. The curve with the parameter $\theta = 0$ corresponds to Planck behaviour.

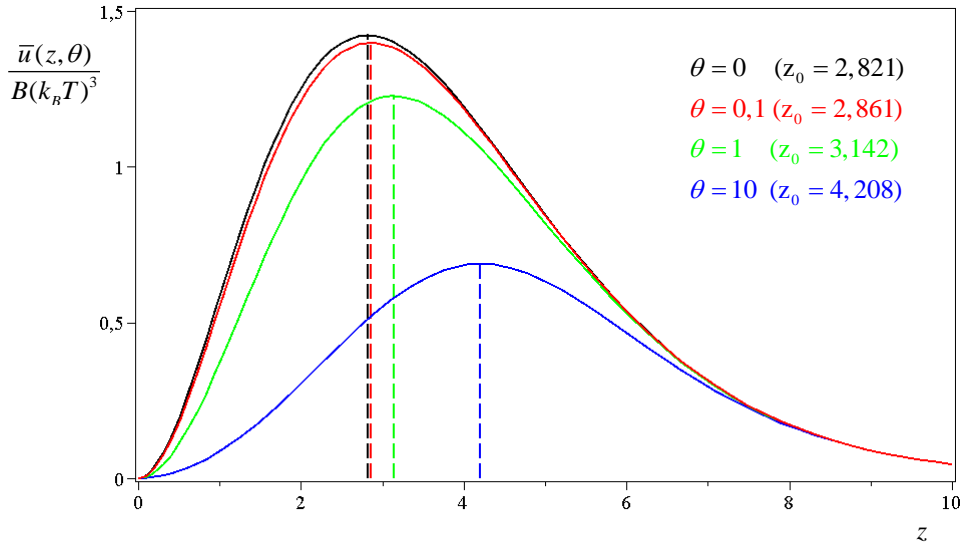

Fig. 2: Spectral energy density $\bar{u}(z,\theta)$ as a function of the photon frequency $z = \hbar\omega / k_B T$ and temperature parameter $\theta = k_B T / E_*$. The curve with the parameter $\theta = 0$ corresponds to Planck behaviour ($B = 1 / \pi^2 \hbar^2 c^3$).

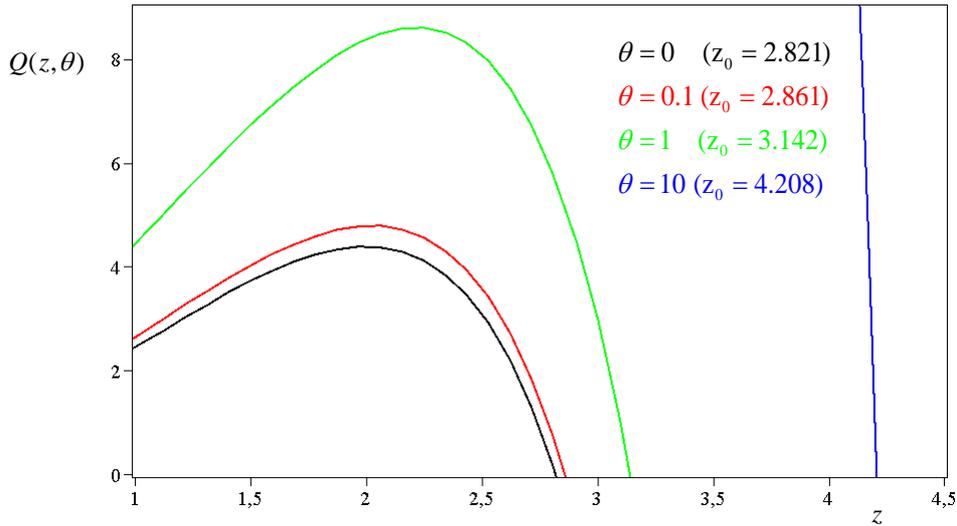

Fig. 3: The zeros $z_0(\theta)$ of the function $Q(z,\theta)$ represent Wien's displacement law in the form $(\hbar\omega)_{max} = z_0 \cdot (k_B T)$. Thereby the value $z_0(\theta = 0) = 2.821$ describes the Planck behaviour.

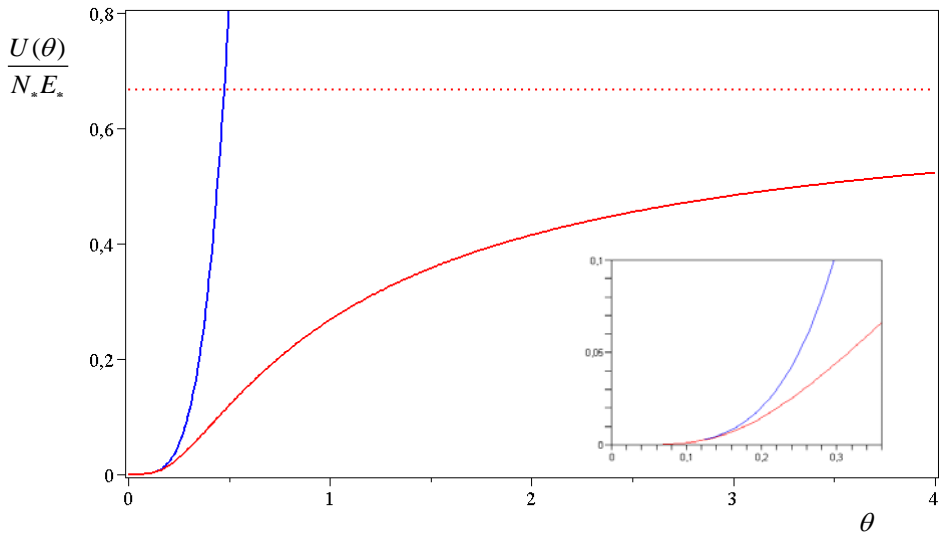

Fig. 4: The internal energy of the photon gas $U(\theta)$ as a function of the temperature parameter $\theta = k_B T / E_*$. The blue curve shows Planck behaviour. In the temperature range up to about $k_B T \sim E_* / 10$, the two curves are almost congruent still, as shown by the inset. For extremely high temperatures $\theta = (k_B T / E_*) \to \infty$, the new, red function curve has a limit $U(\theta) / N_* E_* = 2/3$.

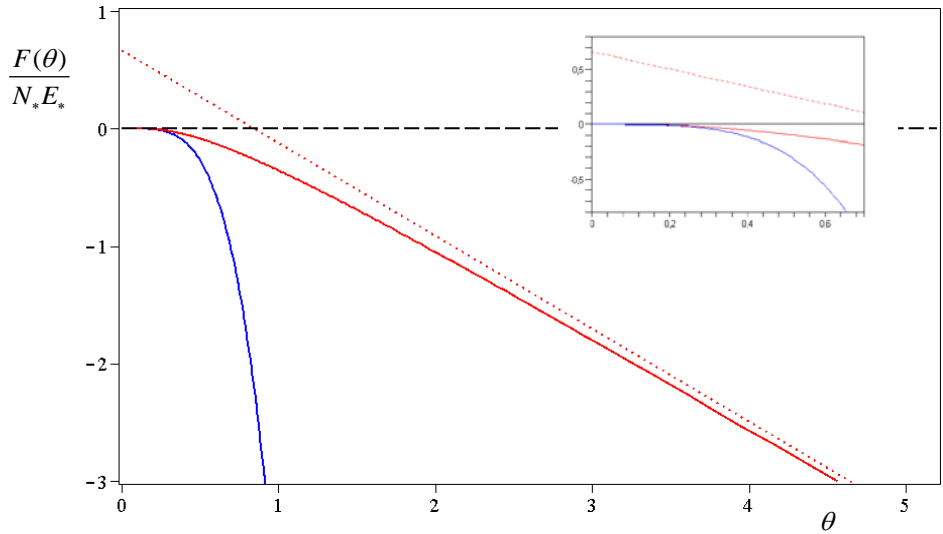

Fig. 5: The free energy of the photon gas $F(\theta)$ as a function of the temperature parameter $\theta = k_B T / E_*$. The blue curve illustrates the Planck behaviour. In the temperature range up to about $k_B T \sim E_* / 10$, the two curves are almost congruent still, as shown by the inset. For extremely high temperatures $\theta = (k_B T / E_*) \to \infty$, the new, red function curve has an asymptote $F(\theta) / N_* E_* \to (2/3) - 2C_\infty \cdot \theta$ (with $2C_\infty \approx 0.79$).

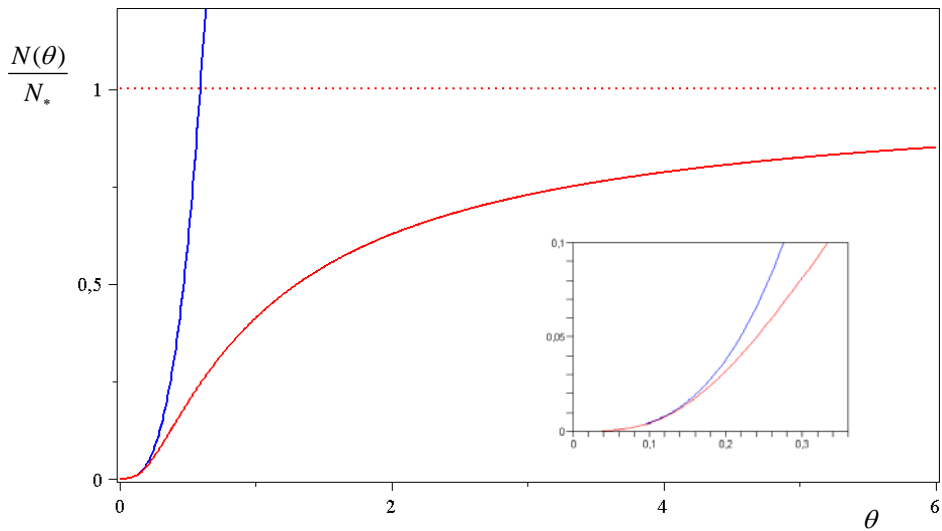

Fig. 6: The total number of particles of the photon gas $N(\theta)$ as a function of the temperature parameter $\theta = k_B T / E_*$. The blue curve shows the Planck behaviour. In the temperature range up to about $k_B T \sim E_* / 10$, the two curves are almost congruent still, as shown by the inset. For extremely high temperatures $\theta = (k_B T / E_*) \to \infty$, the new, red function curve has a limit $N(\theta) / N_* = 1$.

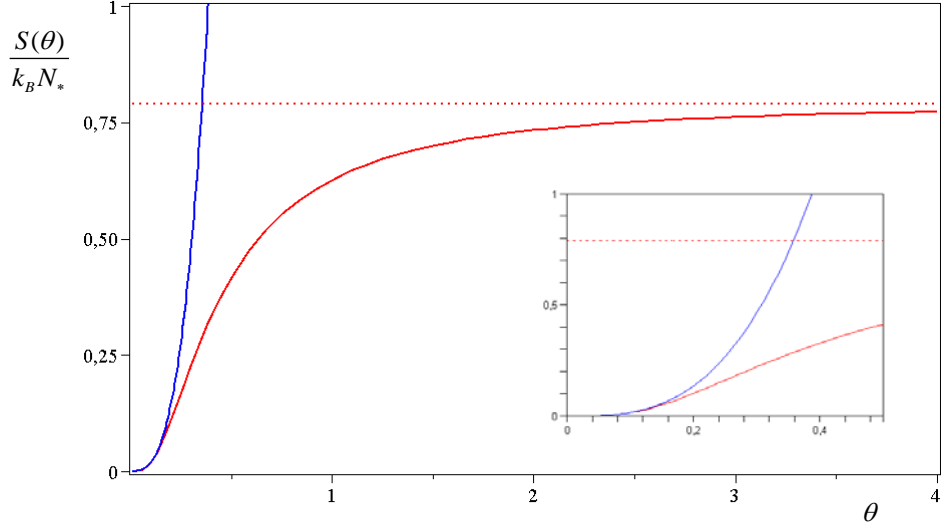

Fig. 7: The entropy of the photon gas $S(\theta)$ as a function of the temperature parameter $\theta = k_B T / E_*$. The blue curve shows the Planck behaviour. In the temperature range up to about $k_B T \sim E_* / 10$, the two curves are almost congruent still, as shown by the inset. For extremely high temperatures $\theta = (k_B T / E_*) \to \infty$, the new, red function curve has a limit $S(\theta) / N_* k_B = 2C_\infty \approx 0.79$.

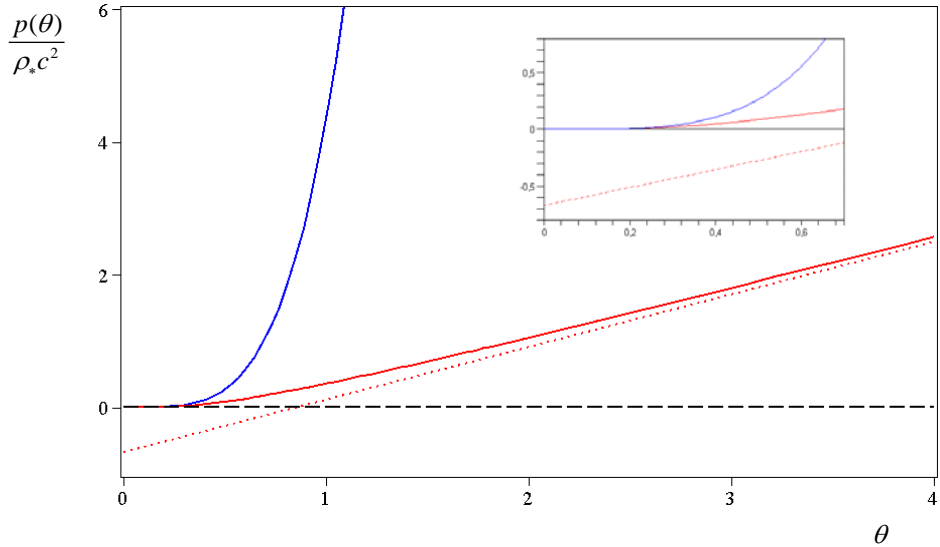

Fig. 8: The pressure of the photon gas $p(\theta)$ as a function of the temperature parameter $\theta = k_B T / E_*$. The blue curve shows the Planck behaviour. In the temperature range up to about $k_B T \sim E_* / 10$, the two curves are almost congruent still, as shown by the inset. For extremely high temperatures $\theta = (k_B T / E_*) \to \infty$, the new, red function curve has an asymptote $p(\theta) / \rho_* c^2 = 2C_\infty \cdot \theta - 2/3$ (with $2C_\infty \approx 0.79$).

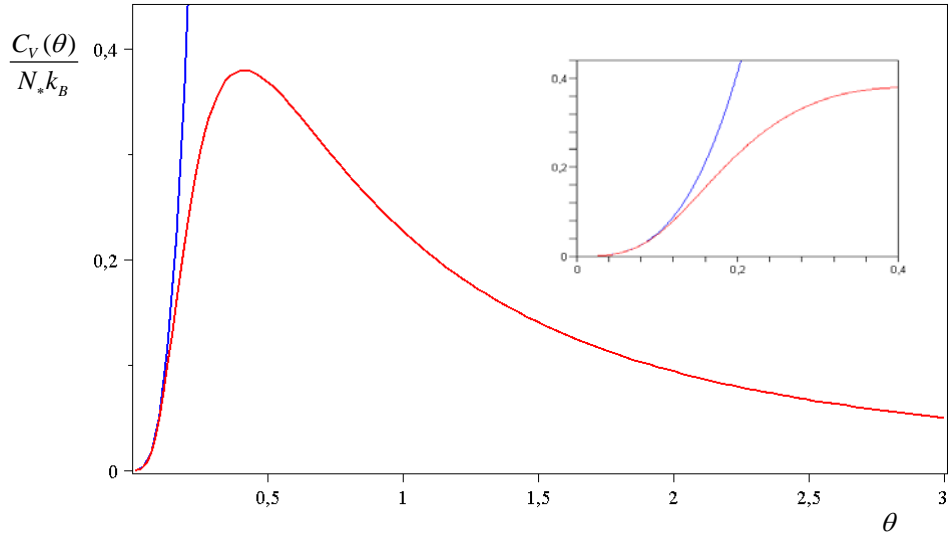

Fig. 9: The specific heat $C_V(\theta)$ as a function of the temperature parameter $\theta = k_B T / E_*$. The blue curve shows the Planck behaviour. In the temperature range up to about $k_B T \sim E_*/10$, the two curves are almost congruent still, as shown by the inset. For extremely high temperatures, the new, red function curve has a maximum at approximately $k_B T \approx E_*/2$.

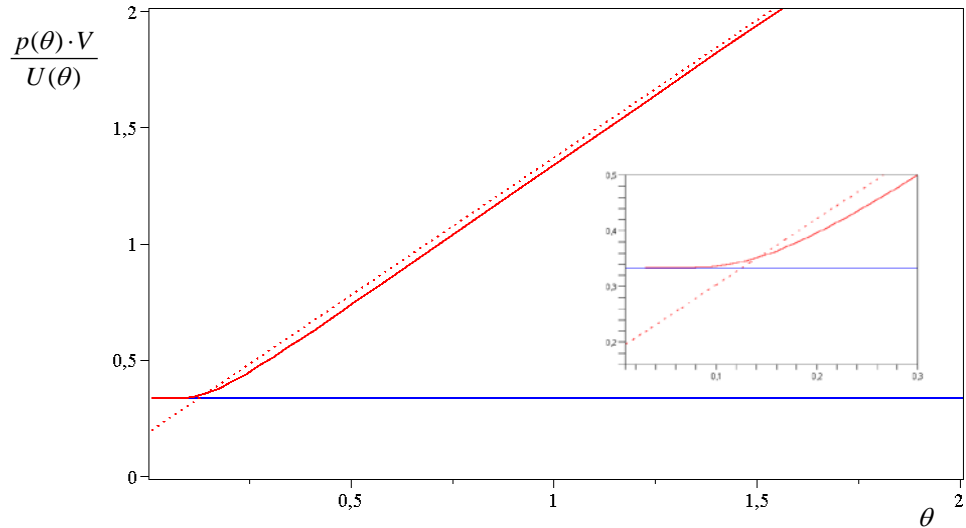

Fig. 10: The ratio $pV/U$ as a function of the temperature parameter $\theta = k_B T / E_*$. The blue curve describes the constant Planck behaviour $pV/U = 1/3$. In the temperature range up to about $k_B T \sim E_*/10$, the two curves are almost congruent still, as shown by the inset. For extremely high temperatures, the new, red function curve shows a linear rise with the asymptote $pV/U = 3C_\infty \cdot \theta + (3C_\infty - 1)$.